\documentclass[reprint,aps,pre,superscriptaddress]{revtex4-1}
\usepackage{amsmath}
\usepackage{amsthm}
\usepackage{amsfonts}
\usepackage{bm}
\usepackage{enumitem}
\usepackage{graphicx}
\graphicspath{{figure/}}
\usepackage{epstopdf}
\usepackage{longtable}
\newcommand{\avg}[1]{\langle {#1} \rangle}
\newcommand{\pderiv}[2]{\frac{\partial {#1}}{\partial {#2}}}
\newcommand{\kb}{k_{\text{B}}}
\let\originalleft\left
\let\originalright\right
\renewcommand{\left}{\mathopen{}\mathclose\bgroup\originalleft}
\renewcommand{\right}{\aftergroup\egroup\originalright}

\newcommand{\dd}{\mathrm{d}}
\newcommand{\ud}{\,\mathrm{d}}

\begin{document}

\title{Population dynamics of driven autocatalytic reactive mixtures}

\author{Hongbo Zhao}
\affiliation{Department of Chemical Engineering}
% \affiliation{Department of Chemical Engineering, Massachusetts Institute of Technology \\77 Massachusetts Avenue, Cambridge, MA 02139}
\author{Martin Z. Bazant}
\affiliation{Department of Chemical Engineering}
\affiliation{Department of Mathematics, Massachusetts Institute of Technology \\77 Massachusetts Avenue, Cambridge, MA 02139}
% \affiliation{Department of Mathematics, Massachusetts Institute of Technology \\77 Massachusetts Avenue, Cambridge, MA 02139}

\date{\today}

\begin{abstract}
  Motivated by the theory of reaction kinetics based on nonequilibrium thermodynamics~\cite{Bazant2013} and the linear stability of driven reaction-diffusion\cite{Bazant2017}, we apply the Fokker-Planck equation to describe the population dynamics of an ensemble of reactive particles in contact with a chemical reservoir. We illustrate the effect of autocatalysis on the population dynamics by comparing systems with identical thermodynamics yet different reaction kinetics. The dynamic phase behavior of the system may be entirely different from what its thermodynamics may suggest. By defining phase separation for a particle ensemble to be when the probability distribution is bimodal, we find that thermodynamic phase separation may be suppressed by autoinhibitory reactions, while autocatalysis enhances phase separation and in some cases induce the ensemble that consists of thermodynamically single-phase systems to segregate into two distinct populations, which we term fictitious phase separation. Asymmetric reaction kinetics also results in qualitatively different population dynamics upon reversing the reaction direction. In the limit of negligible fluctuations, we use method of characteristics and linearization to study the evolution of the standard deviation of concentration as well as the condition for phase separation, in good agreement with the full numerical solution. Applications are discussed to Li-ion batteries and {\it in situ} x-ray diffraction.
\end{abstract}

\maketitle

\section{Introduction}
Population dynamics has been studied widely in many particulate systems, such as crystallization, aerosol dynamics, emulsion, cell culture, etc \cite{Ramkrishna2000,Ramkrishna2014}. A common feature of these systems is the existence of intrinsic instability, either thermodynamic or kinetic in origin, that drives the evolution of the probability distribution. For example, in a multi-phase system, the path toward minimizing interfacial energy causes the mean and the variance of the size of the dispersed phase to increase in time, a well-studied phenomenon called Ostwald ripening\cite{Lifshitz1961}. Kinetically driven examples of population dynamics include cell growth, where the autocatalytic cell division rate can be connected to the population growth rate \cite{Jafarpour2018}, as well as bistability observed in gene regulatory processes\cite{Shu2011}.

Our work is motivated by such nonlinear dynamics in the context of an ensemble of particles that exhibit bistable reaction kinetics, in particular electrochemical intercalation reactions. Dreyer et al.\cite{Dreyer2010, dreyer2011physica, Dreyer2011, Dreyer2015, Guhlke2018} used Fokker-Planck equation to describe the probability distribution of the state of particles in many-particle storage systems such as lithium ion batteries and interconnected rubber balloons, in particular the phase transition due to regions of instability where the chemical potential decreases with lithium concentration and pressure decreases with volume. The model is capable of predicting voltage hysteresis between slow charge and discharge. Other studies\cite{Mielke2012,Dreyer2011} have used similar equations to model mechanical systems as an ensemble of bistable units. Extensive study has also been done on the bistable kinetics of CO oxidation \cite{Bonnefont2017,Crespo-Yapur2013,Johanek2004}, which shows S-shaped negative differential resistance in the current-voltage relationship as well as bistability of surface coverage on Pd nanoparticles. Of particular interest is the population dynamics of coupled microelectrodes \cite{Crespo-Yapur2014}, where sequential activation of CO oxidation upon current ramping is observed.

Emergent behavior arises in population dynamics with nonlinear dynamics. Herrmann et al. \cite{Herrmann2012, Herrmann2014} studied phase transition based on Dreyer's model in the case of constant total current. The coupling of particles through the constraint of total current gives rise to a variety of regimes ranging from Kramers type transition, oscillatory bimodal distribution, to unimodal distribution with increasing total current. Other examples of emergent behavior have also been observed through nonlocal interaction via moments of distribution\cite{Frank2005} or globally coupling mean field parameters such as those in the synchronization of Kuramoto coupled oscillators\cite{Strogatz2000}. Concerted behaviors observed in crowd and biological synchrony \cite{Mirollo1990,Strogatz2005} and recently in Li-ion batteries\cite{Li2018} motivate further theoretical modeling.

The reaction kinetics for lithium intercalation in models above is linear with respect to the thermodynamic driving force, lending it the same kinetic stability conditions as its thermodynamic stability. However, asymmetry between charge and discharge of lithium ion battery is observed in particle ensemble experiments such as chronoamperometry\cite{Bai2013}, as well as single-particle imaging experiments \cite{Li2014,Lim2016}, which have shown that the level of Li concentration heterogeneity is asymmetric between charge and discharge. Such asymmetry cannot be captured by a symmetric linear kinetics.

Bazant\cite{Bazant2017} extended the stability of open driven systems to reactions with general reaction kinetics under a thermodynamically consistent framework. The kinetic stability landscape of the system can be altered by an explicitly state-dependent autocatalytic or autoinhibitory reaction kinetics, showing the suppression of thermodynamic instability in one direction and enhancement in the other, or the creation of instability in thermodynamically stable systems. For a system driven far from equilibrium, we demonstrate that the interplay between reaction kinetics and thermodynamics results in a rich collection of phenomenon observed not only for a single system, but also for a reaction-controlled particle ensemble. State-dependent kinetics has also been explored in velocity dependent or space dependent friction chemotaxis and Van der Pol oscillator \cite{Schweitzer2007}, which leads to undamped limit cycle motion or several stable states in the phase plane. There have also been studies on temporally and spatially dependent friction in activated rate processes using generalized Langevin equation and Fokker-Planck equation \cite{Haynes1995,Haynes1993,Voth1992,Straus1993,Schell1981,Pollak1993}.

The state-dependent reaction kinetics has also been implemented in the population dynamics in realistic electrochemical systems such as lithium iron phosphate (LFP) and graphite, pioneered by Multiphase Porous Electrode Theory (MPET) \cite{Smith2017, Burch2009, Ferguson2014, Ferguson2012}. MPET simulates discrete particles of random sizes in a porous electrode and resolves the spatial dependence including the phase separation within the particles through Allen-Cahn/Cahn-Hilliard reaction model\cite{Bazant2013}. It also incorporates transport equations in the electrolyte. For phase separating materials, at low current, discrete particles are filled stochastically, a phenomenon known as mosaic instability. The results of MPET model have been validated experimentally in a direct observation of lithium concentration in an ensemble of LFP nanoparticles\cite{Li2014}. A statistical model has also been applied to explain the chronoamperometric response of LFP electrodes\cite{Bai2013}. However, only the fractions of untransformed, active and transformed particles are considered, instead of a continuous probability distribution. MPET includes all relevant physics and has shown predictive power in realistic systems but lacks an accurate statistical description as it is limited by the number of particles used in the simulation. Our theory eliminates any spatial dependence and is capable of solving for the probability distribution directly in the reaction-controlled limit.

In section \ref{sec:theory}, we lay out the theory for population dynamics, starting from its mathematical description in terms of the probability distribution in section \ref{sec:single_to_ensemble}. Based on the thermodynamics of an ensemble of particles in section \ref{sec:thermodynamics}, we construct a thermodynamically consistent theory for population dynamics based on generalized Fokker-Planck equation and Langevin equation in section \ref{sec:dynamics}. Section \ref{sec:kinetics} explores a variety of reaction kinetics and studies how their particular functional form impacts the dynamics analytically. In the end, section \ref{sec:applications} applies the theory to LFP, for which experimental data are available, and to a model system that is at the critical point.

\section{Theory}
\label{sec:theory}
\subsection{From single particle dynamics to population dynamics} \label{sec:single_to_ensemble}

We begin by considering a particle that undergoes reactions when it is out of chemical equilibrium with its surrounding,
\begin{equation}
  \text{M}_{\text{res}} \rightarrow \text{M},
\end{equation}
where the single component of interest M reacts with a reservoir with species $\text{M}_{\text{res}}$. The dynamics is described by the change in the concentration of M in the particle,
\begin{equation} \label{eqn:traj_determ}
  \frac{\dd c}{\dd t} = R(c,\mu,\mu_{\text{res}}),
\end{equation}
where $R$ is the reaction rate as a function of concentration $c$, chemical potential of M in the particle $\mu$ and of $\text{M}_{\text{res}}$ in the reservoir $\mu_{\text{res}}$. The reaction rate is zero when the chemical reaction is at equilibrium $\mu = \mu_{\text{res}}$. 

Following Bazant \cite{Bazant2017}, we define the total autocatalytic rate of the reaction to be
\begin{equation}
  s = \pderiv{R}{c} =  \left(\pderiv{R}{c}\right)_\mu + \left(\pderiv{R}{\mu}\right)_c \frac{\dd \mu}{\dd c}.
\end{equation}
which determines whether the concentration is linearly stable ($s<0$) or unstable ($s>0$) to infinitessimal perturbations.  (See Ref. ~\cite{Bazant2017} for a general stability theorem for multicomponent driven reactive mixtures, in terms of the chemical diffusivity and autocatalytic rate tensors.) In most classical theories, the reaction rate is linear in the thermodynamic driving force, or affinity of the reaction~\cite{Kondepudi2014}. More generally, if the reaction rate is linearized near equilibrium, $R = R_0 \Delta \mu$, for small thermodynamic driving force, $\Delta \mu = \mu_{\text{res}}-\mu$ with constant exchange rate prefactor, $R_0$, the autocatalytic rate is given by  $s=- R_0 \, \dd \mu/\dd c$. In this case of near-equilibrium reactions, we arrive at the Duhem-Jougeut Theorem~\cite{Kondepudi2014}, which states that kinetic stability ($s<0$) is equivalent to thermodynamic stability ($\dd \mu/\dd c>0$).   

Far from equilibrium, the situation is very different, and kinetic stability is altered by the explicit dependence of the reaction rate on concentration or chemical potential~\cite{Bazant2017}. The concentration profile tends to be destabilized by a positive ``solo-autocatalytic rate", $\left( \partial R / \partial c \right)_\mu>0$ or, in the case of a thermodynamically stable system, $\dd \mu/\dd c>0$, by a negative``differential reaction resistance" $\left( \partial R / \partial \mu \right)_c< 0$ (as in the case of Marcus kinetics for electron transfer reactions~\cite{Bazant2013}).   The stability of the system is hence determined by both thermodynamics and reaction kinetics. The interplay between the two is the source of inspiration for the following discussion.

The analysis of kinetic stability described here is applicable to lithium intercalation reaction in LFP, a phase-separating electrode material. In particular, using a thermodynamically consistent reaction kinetics, Bazant and coworkers \cite{Bai2011, Cogswell2012, Bazant2013} showed a linearly stable regime in the spinodal region at large enough reaction rate, leading to the suppression of phase separation. The reaction kinetics has been coupled with phase field and mechanical models to predict the spatial pattern of lithium concentration within nanoparticles \cite{Bai2011, Cogswell2012, Cogswell2013, Smith2017, Nadkarni2018}. In the Allen-Cahn reaction (ACR) model \cite{Bazant2013}, the Cahn-Hilliard term $-\kappa \nabla^2 c$ is added to the chemical potential $\mu$ to penalize concentration gradient and allow for a diffuse interface. The assumption behind the ACR model that homogeneous reaction takes place in a particle is shown to be valid when surface reaction dominates while the concentration variation in the direction normal to the surface is negligible.

In the limit of vanishing spatial gradient within particles, the models above for lithium intercalation reduce to Eq. \ref{eqn:traj_determ}.
In the limit of infinite number of reaction-controlled particles that are small enough to assume intraparticle homogeneity, we switch from a Lagrangian description of the discrete particle dynamics to an Eulerian description in terms of a probability distribution,
\begin{equation} \label{eqn:Eulerian}
  \pderiv{f}{t} + \pderiv{\left( fR \right)}{c} = 0,
\end{equation}
where $f(t,c)$ is the probability distribution function (PDF).

\begin{figure}[htb]
  \centering
  \includegraphics[width=\columnwidth]{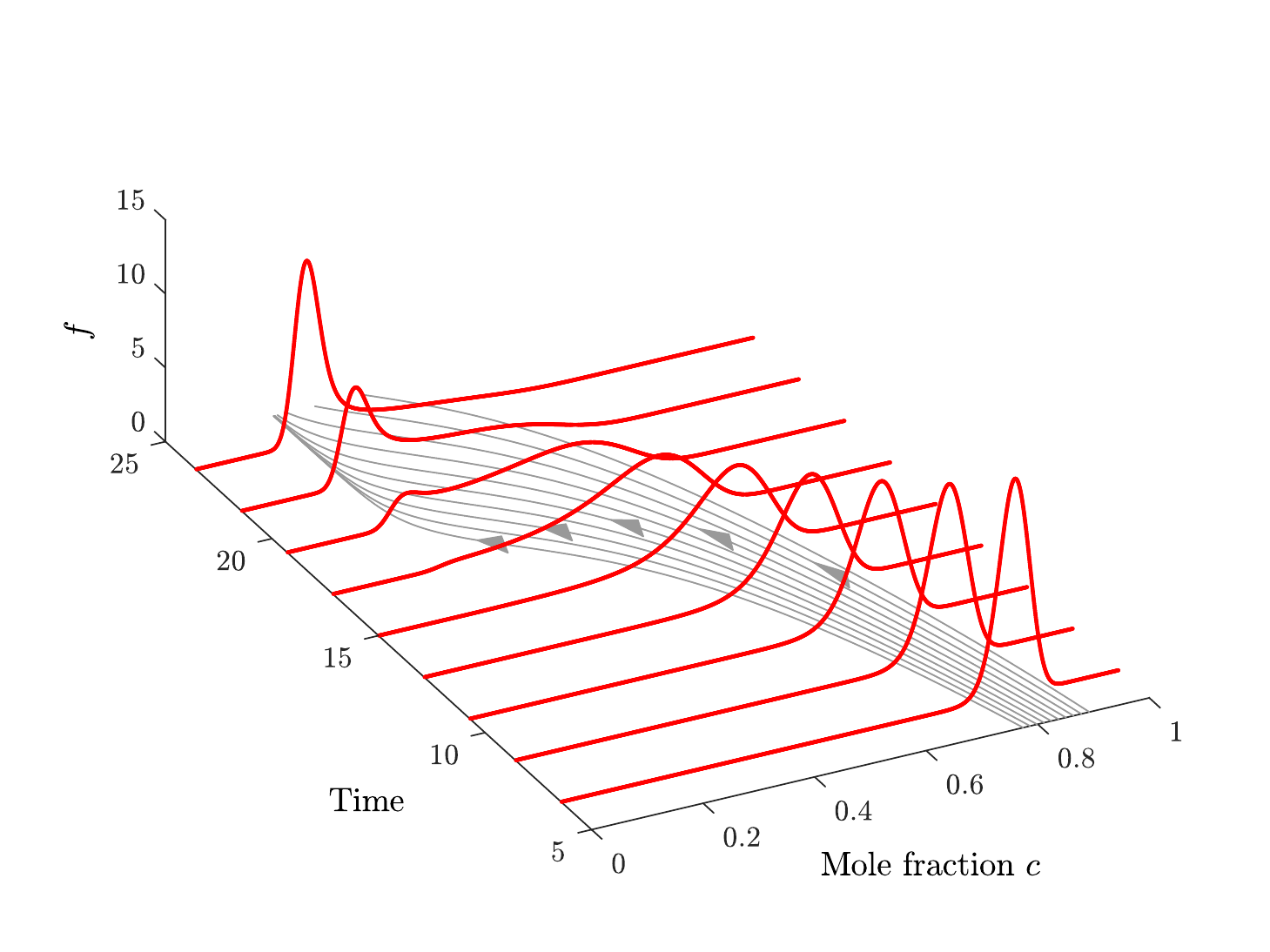}
  \includegraphics[width=\columnwidth]{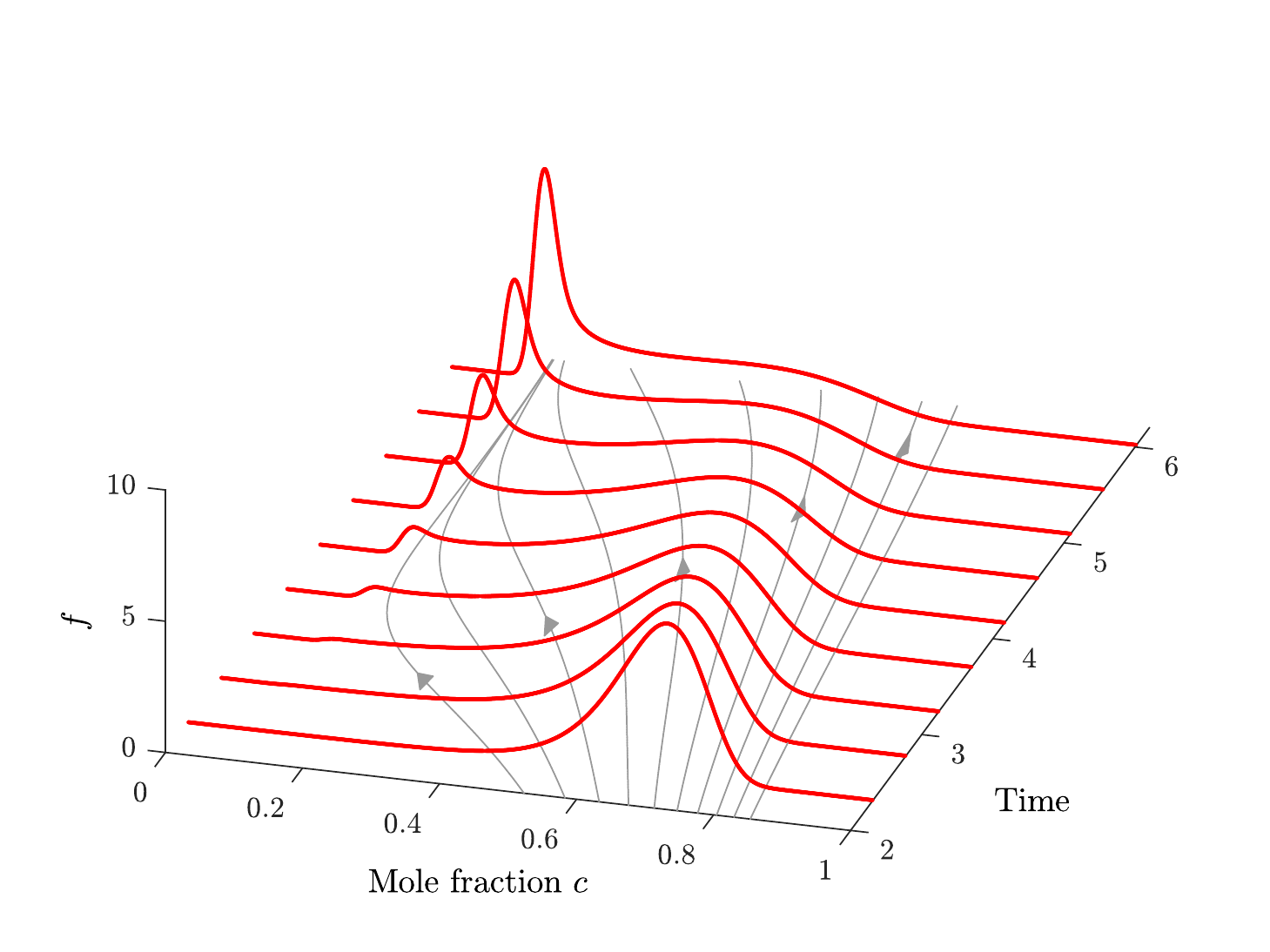}
  \caption{The evolution of probability distribution $f(c,t)$ in time ($D_0 = 10^{-3}$), shown by the red curves. The system undergoes reaction at constant external reservoir chemical potential (left) and constant total reaction rate with time-dependent reservoir chemical potential (right). The gray curves on the horizontal plane are characteristics, or solutions to the deterministic reaction rate equation $\dd c/\dd t = R$ starting from different initial conditions. The distribution widens where characteristics diverge due to autocatalysis and it sharpens where characteristics converge due to autoinhibition. \label{fig:demo_autocat}}
\end{figure}

The characteristics of the linear advection equation in the $c-t$ space is given by the reaction kinetics in Eq. \ref{eqn:traj_determ} \cite{Ramkrishna2000}. When the reaction is autocatalytic, $\partial R/\partial c>0$, characteristics diverge and the distribution expands, whereas when $\partial R/\partial c<0$, characteristics converge and the distribution forms a shock. Fig. \ref{fig:demo_autocat} illustrates the effect of autocatalysis on the evolution of $f(t,c)$ for two cases where a unimodal distribution evolves into a bimodal distribution over time. The evolution of $f(t,c)$ follows the characteristics, or the gray curves, which are solutions to Eq. \ref{eqn:traj_determ}. Due to autocatalysis, particles ahead of others in the reaction direction accelerate and form another peak in the PDF around a concentration removed from the original peak. The formation of peaks in the distribution is analogous to compression shock waves, while broadening distribution is analogous to rarefaction shock waves studied extensively in convection equations. The following sections establish the foundation of this mathematical description in thermodynamics and kinetics.

\subsection{Thermodynamics of a particle ensemble} \label{sec:thermodynamics}
The model above assumes deterministic reaction rate. Now we seek a statistical mechanical description for an ensemble of particles, where thermodynamic fluctuations are important. For simplicity, we consider single-component systems. In the canonical ensemble, the number of M in the system is fixed to be $N$. In the grand canonical ensemble, the system is in a chemical reservoir of $\text{M}_\text{res}$ with chemical potential $\mu_\text{res}$. In table \ref{table:thermo_quant} we define thermodynamic quantities in canonical and grand canonical ensembles as a function of the probability distribution \cite{Chandler1987}. $p_\nu$ is the probability of a particular microstate $\nu$, $p_{\nu_N}=p(\nu|N)$ is the conditional probability of a microstate given $N$ molecules, and $p_N$ is the marginal probability of having $N$ molecules. The probability distribution can be generally in or out of equilibrium. For the canonical ensemble, the subscripts $N$ in the free energy $A_N$, average energy $\avg{E}_N$ and entropy $S_N$ denote the number of molecules specified for the ensemble. We are particularly interested in how the thermodynamic quantities in the grand canonical ensemble are associated with those in the canonical ensemble. For example, using the property of conditional probability $p_{\nu} = p_{\nu_N} p_N$, the grand canonical entropy $S$, the canonical entropy $S_N$ and the entropy due to mole fluctuations $S'$ are related by
\begin{equation}
S = S' + \avg{S_N},
\end{equation}
where $\avg{S_N} = \sum_N{p_N S_N} = -\kb \sum_{\nu}{p_{\nu}\ln{p_{\nu_N}}}$ is the canonical entropy averaged over $N$, and is also known as conditional entropy in information theory.

The free energies of the two ensembles are also related as shown in table \ref{table:thermo_quant}. In particular, if all degrees of freedom except for $N$ is at equilibrium, the grand canonical ensemble free energy is
\begin{equation}
  \Psi = \sum_N{p_N\left( A_N^\text{eq} - \mu_{\text{res}}N + \kb T \ln{p_N} \right)}.
\end{equation}
This may describe the free energy of an ensemble of particles that have reached equilibrium within themselves but not necessarily with the reservoir.

For convenience, we use $A = A_N^\text{eq}$ to denote the canonical ensemble free energy at equilibrium. From thermodynamics, the system is stable when, $\partial^2 A / \partial N^2 > 0$, if $N$ is treated as a continuous variable. An example of instability, which leads to phase separation, can arise in the regular solution model for lattice fluid discussed in section \ref{sec:kinetics}. Since the chemical potential of $M$ within the particle is defined to be $\mu = \partial A/\partial N$, the stability criterion can also be written as $\partial \mu / \partial N>0$. This is not to be confused with the fact that, at equilibrium, $\partial^2 \Psi^{\text{eq}}/\partial \mu_{\text{res}}^2 = - \partial \avg{N}^\text{eq} / \partial \mu_{\text{res}} = - \kb T \sigma_N^2 <0$, where $\sigma_N^2$ is the variance of $N$. Therefore, at equilibrium, the average number of M in an grand canonical ensemble always increases with reservoir chemical potential regardless of $A$. Note that $\sigma_N^2$ at equilibrium is constructed naturally as a result of the entropy $S'$ associated with mole fluctuations.

At equilibrium, the probability distribution satisfies the Boltzmann distribution and minimizes the free energy. Therefore we may define a Lyapunov function to be free energy $\Psi$ minus its minimum $\Psi^{\text{eq}}$, which is always greater than or equal to zero and, as will be shown later, monotonically decreases for systems with a constant $\mu_{\text{res}}$. The Lyapunov function is called relative entropy, or Kullback-Leibler divergence, which originates in information theory to measure the difference between two probability distributions.
\begin{equation}
  \mathcal{L} = \Psi - \Psi^{\text{eq}} = \kb T \sum_{\nu}{p_{\nu} \ln{\frac{p_\nu}{p_\nu^{\text{eq}}}}} > 0.
\end{equation}

\begin{table*}[hbt]
  \caption{Definitions of thermodynamic quantities. $\avg{\cdot}$ denotes averaging over states in the summation that comes immediately before its appearance. $\beta = 1/\kb T$.}
  \label{table:thermo_quant}
\begin{center}
\begin{ruledtabular}%{ccc} \hline \hline
  \begin{tabular}{ccc}
    Ensemble type & Free energy & Entropy \\ \hline
    Canonical & $\begin{array}{rl} A_N &= \sum_{\nu_N}{p_{\nu_N} \left( E_{\nu_N} + \kb T \ln{p_{\nu_N}} \right)}  \\ &= \avg{E}_N - TS_N \end{array}$ & $S_N = -\kb \sum_{\nu_N}{p_{\nu_N}\ln{p_{\nu_N}}}$ \\ \hline
    Grand canonical & $\begin{array}{rl} \Psi &= \sum_{\nu}{p_{\nu}\left( E_{\nu} - \mu_{\text{res}}N_{\nu} + \kb T \ln{p_{\nu}} \right)} \\ &= \avg{E} - \mu_{\text{res}}\avg{N} - TS \\ &=\sum_N{p_N\left( A_N - \mu_{\text{res}}N + \kb T \ln{p_N} \right)} \\ &= \avg{A_N} - \mu_{\text{res}}\avg{N} - TS'\end{array}$ & $\begin{array}{rl} S &= -\kb \sum_{\nu}{p_{\nu}\ln{p_{\nu}}} \\ S' &= -\kb \sum_{N}{p_N \ln{p_N}} \end{array}$ \vspace{1pt} \\ \hline

    Ensemble type & \multicolumn{2}{c}{Equilibrium distribution} \\ \hline
    Canonical & \multicolumn{2}{c}{$p_{\nu_N}^{\text{eq}} = \exp{\left[ -\beta \left( E_{\nu_N} + A_N^{\text{eq}} \right) \right]}$} \\ \hline
    Grand canonical & \multicolumn{2}{c}{$p_{\nu}^{\text{eq}}= \exp{\left[ -\beta \left( E_{\nu} + \mu_{\text{res}} N_{\nu} + \Psi^{\text{eq}} \right) \right]},$ \hspace{1cm} $p_N^{\text{eq}} = \exp{\left[-\beta \left( A_N^{\text{eq}} + \mu_{\text{res}}N + \Psi^{\text{eq}} \right) \right]}$} \\
% \hline \hline
  \end{tabular}
\end{ruledtabular}
\end{center}
\end{table*}

\subsection{Dynamics}
\label{sec:dynamics}
In the classical limit, we use a general continuous variable $\bf{x}$ that maps the microstate of the system $\nu$ to $\mathbb{R}^n$ and use $p$ to denote the probability density function. In the grand canonical ensemble, the driving force is the gradient of the potential in the phase space. The potential is defined by the variational derivative $\delta \Psi / \delta p = E - \mu_{\text{res}}N + \kb T \ln{p}$. Note that equilibrium corresponds to $\delta \Psi / \delta p$ being a constant that satisfies the normalization condition for $p$. Assuming the flux in the phase space is proportional to the driving force, we generalize the Fokker-Planck equation, which can be regarded as Wasserstein gradient flow\cite{Risken1989, Herrmann2012, Jordan1998},
\begin{equation} \label{eqn:FP_general}
  \frac{\partial p}{\partial t} + \nabla \cdot \mathbf{J} = 0,
\end{equation}
where
\begin{equation}
  \mathbf{J} = -\mathbf{K} p \nabla \frac{\delta \Psi}{\delta p} =-\mathbf{K} \kb T p \nabla \ln{\frac{p}{p^{\text{eq}}}},
\end{equation}
where $\bf{K}$ is the mobility tensor.

The following second law of thermodynamics is well known\cite{Sekimoto2010, Reguera2005, Rao2016, Seifert2005, Seifert2012, VandenBroeck2010, Frank2005},
\begin{equation}
  \frac{\dd \Psi}{\dd t} = -\int_{\mathbb{R}^n}{\frac{\mathbf{J}^{\dagger} \mathbf{K}^{-1} \mathbf{J}}{p} \ud \mathbf{x}} + \int_{\Omega}{p \frac{\dd}{\dd t}\left( E - \mu_{\text{res}}N \right) \ud \mathbf{x}},
\end{equation}
where the first term is the entropy production rate $-T\dot{S}_i$ and the second term is the work of external driving force $\dot{W}_d$. When the system is driven chemically,
\begin{equation}
  \dot{W}_d = - \dot{\mu}_{\text{res}} \langle N \rangle.
\end{equation}
The second law of thermodynamics requires that $\dot{S}_i>0$, which means $\mathbf{K}$ is positive definite. When the system is nondriven, $\Psi-\Psi^{\text{eq}}$ decreases monotonically, hence it is a Lyapunov function.

If the relaxation to equilibrium in terms of its internal degrees of freedom $\nu_N$ is much faster than the exchange of matter with the reservoir, which may take place via chemical reactions, we may assume all internal degrees of freedom are at equilibrium, that is, $p_{\nu}=p_{\nu_N}^{\text{eq}}p_N$. In this case, we can show that the marginal probability in terms of $N$ can be described by the following equation,
\begin{equation} \label{eqn:FP_general_N}
  \frac{\partial p_N}{\partial t} + \frac{\partial J_N}{\partial N} = 0,
\end{equation}
where
\begin{align}
  J_N &= - \left( \mathbf{K} \right)_{N,N} p_N  \frac{\partial}{\partial N} \left( \frac{\delta \Psi}{\delta p_N} \right), \\
  \frac{\delta \Psi}{\delta p_N} &= A_N^{\text{eq}} - \mu_{\text{res}}N +\kb T \ln{p_N}.
\end{align}
It is convenient to nondimensionalize the number of molecules $N$ in the system by the its maximum $N_t$, which is well defined for intercalation systems. Assuming that $N_t$ is fixed, the equation can be nondimensionalized via the mole fraction, $c=N/N_t$, the PDF in terms of $c$, $f=P_N N_t$, and prefactor $k={(\mathbf{K})_{N,N}}/N_t$. In contrast to linear kinetics, $k$ can be a function of $c$, $\mu$, and $\mu_{\text{res}}$, and we do not make any assumption about the functional form other than $k>0$.
\begin{equation} \label{eqn:FP_c}
  \frac{\partial f}{\partial t} + \frac{\partial \left( \left( \mu_{\text{res}} - \mu \right) kf \right)}{\partial c} = \frac{\kb T}{N_t} \frac{\partial}{\partial c}\left( k \frac{\partial f}{\partial c} \right),
\end{equation}
where $\mu=\partial A_N^{\text{eq}}/\partial N$. Alternatively,
\begin{equation*}
    \frac{\partial f}{\partial t} + \frac{\partial j}{\partial c} =0,
  \end{equation*}
where the flux $j$ is,
\begin{equation}
  j= \left( \left(\mu_{\text{res}}-\mu \right) k - k\frac{\kb T}{N_t} \frac{\partial}{\partial c} \right)f = k \kb T f^\text{eq} \frac{\partial}{\partial c}\left( \frac{f}{f^\text{eq}} \right),
\end{equation}
where $f^\text{eq}$ is the equilibrium distribution, given by Eq. \ref{eqn:eqm_dist}. Defining the maximum number of molecules $N_t$ restricts the fraction $c \in [0,1]$. Therefore we impose the no-flux boundary condition $j=0$ at $c=0$ and 1. To be compatible, we require the chemical potential $\mu \to \infty$ as $c\to 1$ and $\mu \to -\infty$ as $c\to 0$ and is undefined outside (0,1).

Following the nondimensionalization, we define the following free energy in the macroscopic sense
\begin{equation} \label{eqn:free_energy}
  \Psi_m = \int_0^1{\left( a -\mu_{\text{res}}c + \frac{\kb T}{N_t} \ln{f} \right) f \ud c},
\end{equation}
where $a=A/N_t$. Following the analysis of the general case (Eq. \ref{eqn:FP_general}), $\Psi_m$ is the Lyapunov function for Eq. \ref{eqn:FP_c}. Therefore if not chemically driven, or at constant $\mu_{\text{res}}$, the system is Lyapunov stable. In fact, it can also be shown that the eigenvalues of the operator $\left( \left(\mu_{\text{res}}-\mu \right)k - k\kb T N_t^{-1} \, \partial/\partial c \right)$ are nonnegative, while the eigenfunction corresponding to the zero eigenvalue is the equilibrium distribution \cite{Risken1989}. In other words, the equilibrium distribution corresponds to $\partial j/\partial c=0$, or $j=0$, based on the boundary condition,
\begin{equation} \label{eqn:eqm_dist}
f^{\text{eq}}(c)=e^{- \beta N_t \left( a - \mu_{\text{res}}c - \Psi_m^{\text{eq}} \right)},
\end{equation}
where $\Psi^{\text{eq}}_m$ is the potential defined in Eq. \ref{eqn:free_energy} when at equilibrium and is the constant that normalizes the distribution.

By comparing Eqs. \ref{eqn:Eulerian} and \ref{eqn:FP_c}, we find that the reaction rate is $R=\left( \mu_{\text{res}} - \mu \right)k$. In the limit of large systems, $N_t \to \infty$, Fokker-Planck equation (Eq. \ref{eqn:FP_c}) reduces to Eq. \ref{eqn:Eulerian}.
Now we are able to correlate mole fluctuations with the thermodynamic quantities in the diffusive term in the Fokker-Planck equation and extend the deterministic dynamics, or Eq. \ref{eqn:traj_determ}, to the following Langevin equation, which is equivalent to the Fokker-Planck equation.
\begin{equation}
  \frac{\dd c}{\dd t} = R + \sqrt{2 \frac{\kb T k}{N_t}} \xi(t),
\end{equation}
where the Langevin noise $\xi(t)$ satisfies zero average $\langle \xi(t) \rangle=0$ and is uncorrelated in time $\langle \xi(t) \xi(t') \rangle = \delta(t-t')$. Eq. \ref{eqn:FP_c} can also be written as,
\begin{equation}
  \frac{\partial f}{\partial t} + \frac{\partial \left( fR \right)}{\partial c} = \frac{\partial}{\partial c} \left( D \frac{\partial f}{\partial c} \right),
\end{equation}
where $D = \kb T k/N_t$ is an expression of the Einstein relation or Fluctuation-Dissipation Theorem for concentration fluctuations across the particle ensemble. The noise intensity decreases as the size of the particle increases.

In section \ref{sec:single_to_ensemble}, we mentioned that when $\partial R/\partial c > 0$, characteristics diverge and the probability distribution expands in time, and vice versa. This statement can be made more quantitative in the case of non-negligible fluctuations by considering the evolution of the variance of the distribution.
Herrmann \cite{Herrmann2012} analyzed the widening of the distribution in the spinodal region for a short time by linearizing the reaction $R$ with respect to $c$, which specifically entails linearizing the chemical potential $\mu$ since the prefactor $k$ is assumed to be constant. Therefore, in this case, the widening of the distribution is determined by $\partial \mu/\partial c$. The analysis requires that the distribution be sharply peaked.

\subsection{ Evolution of the concentration variance }
With the same assumption, we extend the analysis to a general reaction rate expression and approximate the evolution of the concentration variance, $\sigma_c^2=\int_0^1{(c-c_0)^2f(c) \ud c}$.  The distribution $f(t,c)$ satisfies the following Fokker-Planck equation
\begin{equation}
  \frac{\partial f}{\partial t} + \frac{\partial j}{\partial c} =0,
\end{equation}
where $j= fR - \kb T N_t^{-1} k \partial f/\partial c$.
Denote the mean $c_0=\langle c \rangle$, and let us calculate the evolution of the concentration variance
\begin{multline}
  \frac{\dd \sigma_c^2}{\dd t}=\frac{\dd}{\dd t}\left( \int_0^1{(c-c_0)^2 f \ud c} \right) \\ = \int_0^1{(c-c_0)^2 \frac{\partial f}{\partial t} \ud c} = 2\int_0^1{j(c-c_0) \ud c}.
\end{multline}
Expand the current $j$ about $c_0$ and use superscript $(m)$ to denote $m^\text{th}$ derivative with respect to $c$,
\begin{multline}
  j = f\left( \sum_{m=0}^\infty{R^{(m)}(c_0)\frac{\left(c-c_0\right)^m}{m!}}\right) \\ - \frac{\kb T}{N_t} \left( \sum_{m=0}^\infty{k^{(m)}(c_0)\frac{\left(c-c_0\right)^m}{m!}} \right) \frac{\partial f}{\partial c},
\end{multline}
Therefore
\begin{multline}
  \frac{\dd \sigma_c^2}{\dd t}=2\sum_{m=0}^\infty{R^{(m)}(c_0)\frac{\langle \left(c-c_0\right)^{m+1} \rangle}{m!}} \\ + 2\frac{\kb T}{N_t} \sum_{m=0}^\infty{k_0^{(m)}\frac{m+1}{m!} \langle \left(c-c_0\right)^m \rangle}.
\end{multline}
If $f$ is concentrated around $c_0$, higher order moments in the expression above decay very quickly and are negligible. Therefore we retain only moments up to $\langle (c-c_0)^2 \rangle = \sigma_c^2$. In the limit of sharply-peaked distribution, $\kb T/N_t$ and $\sigma_c^2$ are both small, we arrive at the key result,
\begin{equation} \label{eqn:variance}
  \frac{\dd {\sigma}^2_c}{\dd t} = 2\left( \left. \frac{\partial R}{\partial c}\right|_{c_0} \sigma^2_c + \frac{\kb T}{N_t} k_0 \right),
\end{equation}
where $k_0 = k(c_0)$, and $c_0=\langle c \rangle$ is the average mole fraction.  

For autoinhibitory reactions, $\partial R/\partial c<0$, the variance tends to decrease in time while fluctuations work against it, and the particle ensemble stays homogeneous. For autocatalytic reactions, the distribution widens and the particle ensemble becomes more heterogeneous. If the equilibrium distribution is unimodal with a sharp peak, it can be approximated by a Gaussian function by expanding the free energy $a$ in terms of $c$ to second order,
\begin{equation}
f^{\text{eq}}(c) \approx e^{-\beta N_t \mu'(c_0) (c-c_0)^2 /2} e^{-\beta N_t \left( a(c_0)- \mu(c_0)c_0 - \Psi_m^{\text{eq}} \right)},
\end{equation}
where, $\mu_{\text{res}} = \mu(c_0)$ and $\sigma_c^2 = \kb T/\left( N_t \mu'(c_0) \right)$. The same conclusion is obtained by setting Eq. \ref{eqn:variance} to zero, since $\partial R / \partial c = -k_0 \partial \mu / \partial c + \left( \mu_{\text{res}} - \mu \right) \partial k/\partial c$. When the system does not phase separate thermodynamically ($\partial \mu/\partial c>0$), the variance can become greater or smaller than the equilibrium variance when there is a net reaction, depending on the sign of  $\partial k/\partial c$ and the direction of the net reaction. It is the focus of this paper to explore the effect of $k(c)$ on population dynamics. The next section describes how different $k(c)$ models predict dramatically different variance evolution.

In the limit of negligible fluctuations, $N_t \to \infty$, Eq. \ref{eqn:variance} gives the evolution of variance over time,
\begin{equation} \label{eqn:var_no_noise}
  \sigma^2_c(t) = \sigma^2_c(t=0) \exp{\left( 2\int_{0}^{t}{\frac{\partial R}{\partial c} \ud t} \right)}.
\end{equation}
In the same limit, with the initial condition $f(c_0, t=0)=f_0(c_0)$, using method of characteristics, the solution \cite{Ramkrishna2000} to the Fokker-Planck equation without the diffusive term (Eq. \ref{eqn:Eulerian}) is
\begin{equation} \label{eqn:f_no_noise}
  f(c\left(t,c_0 \right),t)=f_0(c_0) \exp{\left( \int_0^t{-\frac{\partial R}{\partial c}\left( c\left(t,c_0 \right),t \right) dt} \right)},
\end{equation}
where $c\left(t,c_0 \right)$ follows the characteristics and is the solution to Eq. \ref{eqn:traj_determ} with the initial condition $c(t)=c_0$.
We see that in the limit of sharply-peaked distribution, by taking the variance on both sides of Eq. \ref{eqn:f_no_noise}, Eq. \ref{eqn:var_no_noise} is recovered.
If the reaction rate does not explicitly depend on  $t$, then
\begin{equation}
  f(c\left(t,c_0 \right),t) = f_0(c_0)\frac{R(c_0)}{R(c\left(t,c_0 \right),t)}.
\end{equation}

\subsection{Electrochemical reactions} \label{sec:kinetics}
Electrochemical systems offer the most physically realizable benchmark for the population dynamics model, since in experiments the control of current or voltage corresponds to the control of the total reaction rate or the chemical potential of the reservoir $\mu_{\text{res}}$, respectively. In this section, we focus on the effect of reaction kinetics in the case of constant total reaction rate -- one of the most commonly used protocol in electrochemistry. The total reaction rate is given by
\begin{equation} \label{eqn:total_rate}
  R_{\text{total}} = \frac{\dd \avg{c}}{\dd t} = \int_0^1{c \frac{\partial f}{\partial t} \ud c} = \int_0^1{j \ud c}.
\end{equation}
In the limit of a sharply-peaked distribution, the center of the peak $c_0$ is equal to the average concentration $\avg{c}$ and shifts linearly in time. The total reaction rate is equal to the reaction rate at $c_0$, that is, $\dd c_0/ \dd t = R(c_0)$. Therefore, the evolution of variance in Eq. \ref{eqn:variance} can also be written as,
\begin{equation}
  \frac{\dd {\sigma}^2_c}{\dd c_0} = 2\left( \left. \frac{s}{R} \right|_{c_0} \sigma^2_c + \frac{\kb T}{N_t} \frac{1}{\mu_{\text{res}}-\mu} \right).
\end{equation}
In the limit of negligible fluctuation or infinitely large reaction rate, when the thermodynamic driving force $\mu_{\text{res}}-\mu$ is large, the growth of the standard deviation of the distribution is given by
\begin{equation} \label{eqn:K}
  K = \ln{\left(\frac{\sigma_c}{\sigma_{c_0}} \right)} = \int_{c_0}^c{\frac{s}{R} \ud c}.
\end{equation}

In this section, we consider the following simple reaction kinetics,
\begin{equation}
  R = R_0(c) g(\mu_{\text{res}}-\mu),
\end{equation}
where $g(0) = 0$, which includes important cases of electrochemical reactions~\cite{Bazant2013,Bazant2017}. The corresponding autocatalytic rate is
\begin{equation}
  s = \frac{\partial R}{\partial c} = R_0'g - R_0 g' \mu' = R\left( \frac{R_0'}{R_0} - \frac{g'}{g}\mu'\right).
\end{equation}
With this particular form of reaction kinetics, we have 
\begin{equation}
  K = \left. \ln{R_0} \right|_{c_0}^{c} - \int_{c_0}^{c}{\frac{g'}{g}\mu' \ud c}.
\end{equation}

Here we focus on the explicit dependence of reaction rate on concentration, in particular how the exchange current, $R_0(c)$, affects the stability of the particle ensemble by electroautocatalysis~\cite{Bazant2017}. Hence we discuss the case of positive differential resistance $g'>0$, that is, the reaction rate increases with the driving force, which preserves the contribution of the thermodynamic stability to the overall stability. Although not the focus of the paper, Marcus theory for outer-sphere electron transfer, for example, exhibits an inverted region, or negative differential resistance\cite{Bazant2013,Bazant2017}, which can destabilize a thermodynamically stable system and make the reaction autocatalytic. Reaction kinetics of this form can be derived from transition state theory. For example, the famous Butler-Volmer equation has the form above\cite{Bazant2013,Newman2012}, where $R_0$ is called the exchange current and
\begin{equation}
  g(x) = e^{\alpha x} - e^{-(1-\alpha) x},
\end{equation}
where $\alpha$ is the charge transfer coefficient. In the case of symmetric Butler-Volmer kinetics, $\alpha=0.5$, $g'=\sqrt{\left( g/2 \right)^2+1}$. The standard deviation grows in regions where $s>0$, or $\left(\ln{R_0}\right)'>\sqrt{1/4+\left(R_0 / R\right)^2}\mu' \text{sgn}(R)$, as shown by Bazant in Ref. \cite{Bazant2017}.
With Butler-Volmer kinetics, when the driving force $\mu_{\text{res}}-\mu$ is small, $g(x) \to x$, and the reaction rate is slow. Therefore the second term in Eq. \ref{eqn:K} dominates and whether the standard deviation increases or decreases is mainly determined by the thermodynamic stability,
\begin{equation}
  K \approx -\frac{1}{R} \int_{c_0}^{c}{R_0 \mu' \ud c}.
\end{equation}
At high total reaction rate, both the exchange current and thermodynamics determine the evolution of standard deviation.
\begin{equation} \label{eqn:K_inf_current}
  K \approx \left. \ln{R_0} \right|_{c_0}^{c} -
  \begin{cases}
    \alpha \left. \mu \right|_{c_0}^{c} & \text{if } R>0 \\
    -(1-\alpha) \left. \mu \right|_{c_0}^{c} & \text{if } R<0
  \end{cases}.
\end{equation}

The exchange current of the reaction kinetics is crucial in determining the stability of the system\cite{Bazant2017, Lim2016}. The asymmetry in exchange current leads to asymmetric behavior between forward and backward reactions. In the case of LFP, a skewed exchange current results in more uniform concentration profiles during intercalation while greater inhomogeneity during de-intercalation.
To illustrate how the exchange current alters the kinetic stability, we choose regular solution as the thermodynamic model.
\begin{equation} \label{eqn:regular_solution}
  \frac{\mu}{\kb T} = \ln{\frac{c}{1-c}} + \Omega (1-2c),
\end{equation}
where the first term arises from the entropy of mixing and the second term comes from enthalpic interaction. When $\Omega>2$, there exists a spinodal gap and the system phase separates into two phases with low and high concentrations.

\begin{figure}[htb]
  \centering
  \includegraphics[width=\columnwidth]{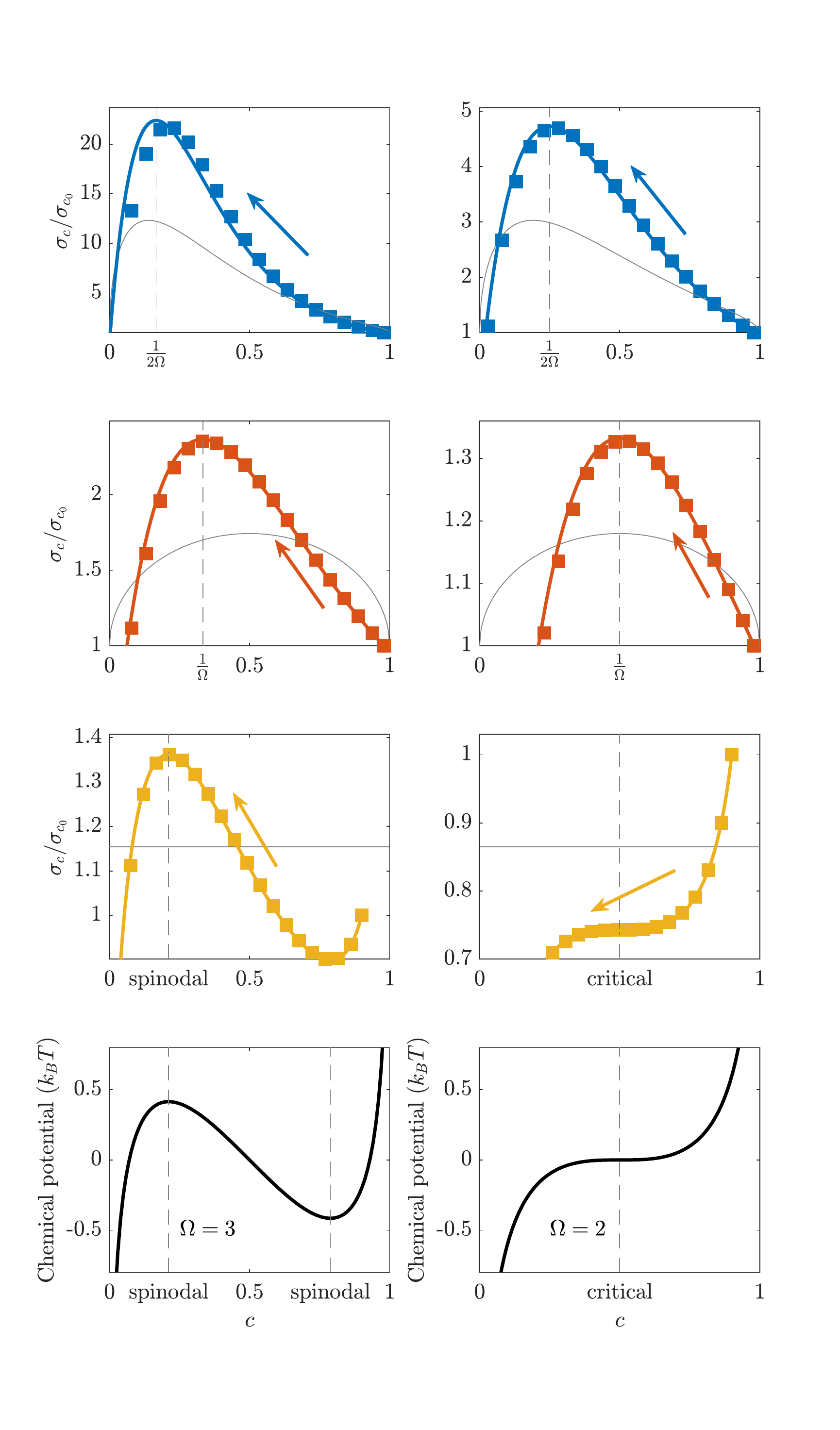}
  \caption{Evolution of the standard deviation of the distribution at infinitely large negative reaction rate and zero fluctuation. The colored solid curves plot the analytical expression (Eq. \ref{eqn:K_inf_current}). Squares represent simulations at sufficiently large negative currents. The location of maximum standard deviation is marked by dashed lines and are predicted in table \ref{table:maxsigma}. Each row corresponds to a different exchange current shown in gray solid curves. Regular solution is used as the thermodynamic model. The left column corresponds to $\Omega = 3$, a phase separating system. The right column corresponds to $\Omega = 2$, which is at the critical point. The arrows indicate the direction of reaction as well as the autocatalytic effect in the case of increasing standard deviation. \label{fig:demo_variance}}
\end{figure}

Table \ref{table:maxsigma} lists, in the case of symmetric Butler-Volmer kinetics ($\alpha=0.5$), the maximum standard deviation ratio at infinitely fast reaction rate for three types of exchange current, which is attained in the interval $[c_1,c_2]$ in which $\partial R/\partial c \geq 0$. The starting and ending concentrations $c_1$ and $c_2$ are also listed in the table. See section \ref{sec:LFP} below for a plot and discussion on the region of instability $\partial R/\partial c \geq 0$ at arbitrary reaction rates. The first two exchange currents are symmetric and thus exhibit symmetric behavior for forward and backward reactions. Constant exchange current does not alter the thermodynamic stability -- the kinetic instability ($s>0$) grows exactly within the spinodal region, which exists when $\Omega>2$. Fig. \ref{fig:demo_variance} shows that evolution of the standard deviation at infinitely large reaction rate. The third row of the figure shows that the evolution for the constant exchange current follows the trend of the chemical potential. For the second exchange current $\sqrt{c(1-c)}$, which is frequently used in electrochemistry, the kinetic instability grows within $[0,1-1/\Omega]$ for forward reaction and $[1/\Omega,1]$ for backward reaction. The instability region exists when $\Omega>1$. Since at low enough reaction rate the system always follows thermodynamic stability, when $1<\Omega<2$, it only becomes linearly unstable at high reaction rate. As shown in the second row of figure \ref{fig:demo_variance}, the second exchange current results in a larger maximum standard deviation ratio as well as region of instability than the constant exchange current.

As Bazant has shown \cite{Bazant2017}, the third exchange current is predominantly autocatalytic in the backward direction due to the asymmetrically increasing exchange current in that direction. When the direction is reversed, the autoinhibition results in greater homogeneity for the forward reaction. In particular, at infinitely large backward reaction rate, the system is kinetically unstable within the concentration range $[1/2\Omega,1]$, which exists when $\Omega>1/2$. At infinitely large forward reaction rate, the system remains stable regardless of the value of $\Omega$, behaving entirely differently from the symmetric exchange currents. Fig. \ref{fig:demo_variance} shows that, in the backward direction, the amplification of standard deviation and the length of the instability interval increases in the order of constant, symmetric $\sqrt{c(1-c)}$ and asymmetric $(1-c)e^{\mu/2}$ exchange current, the reason for the latter being $1/2\Omega<\left( 1-\sqrt{1-2/\Omega} \right)/2 < 1/\Omega$. Fig. \ref{fig:demo_variance} also shows the agreement between the analytical expression for the evolution of standard deviation from Eq. \ref{eqn:K_inf_current} and numerical simulation for the same exchange currents. The approximation captures the magnitude of maximum standard deviation and where maximum standard deviation is attained.

\begin{table*}[htb]
  \caption{Maximum standard deviation ratio with the starting and ending average fraction for three different exchange current.}
  \label{table:maxsigma}
\begin{center}
\begin{ruledtabular}%{c|c|c} \hline \hline
  \begin{tabular}{cccc}
    Exchange current & $1 \; (\Omega>2)$ & $\sqrt{c(1-c)} \; (\Omega>1)$ & $(1-c)e^{\mu/2} \; (\Omega>1/2)$  \\ \hline
    $\max_{c_1,c_2}{\frac{\sigma_{c_1}}{\sigma_{c_2}}}$ & $\frac{1-\sqrt{1-2/\Omega}}{1+\sqrt{1-2/\Omega}} \exp{\left( \Omega \sqrt{1-2/\Omega} \right)}$ & $\frac{1}{\Omega} e^{\Omega-1}$ & $\frac{1}{2\Omega} e^{2\Omega -1}$ \\
    $c_1$ & $\frac{1 \mp \sqrt{1-2/\Omega}}{2}$ & $0$ or $1$ & $1$ \\
    $c_2$ & $\frac{1 \pm \sqrt{1-2/\Omega}}{2}$ & $1-\frac{1}{\Omega}$ or $\frac{1}{\Omega}$ & $\frac{1}{2\Omega}$ \\
  \end{tabular}
\end{ruledtabular}
\end{center}
\end{table*}

When the variance $\sigma_c^2$ gets large, the assumption of sharply-peaked distribution no longer holds. For example, when the system is driven toward high concentration, autocatalytic effect may cause some particles to accelerate to much higher concentration than average. However, the highest concentration a particle can reach is the equilibrium concentration corresponding to $\mu_\text{res}$. The stabilizing effect of thermodynamics when the concentration is near equilibrium ($\partial R/\partial c>0$) causes those particles to form another peak in the distribution. We define the resulting bimodal distribution to be phase separation at the ensemble level.

The key result of this work is to show that phase separation can arise as a result of both kinetic and thermodynamic effects, and mathematically, it is a result of diverging and converging characteristics. Based on the analysis, we may approximate the initiation of phase separation as when the estimated variance $\sigma_c = \sigma_{c_0}e^K$ exceeds a certain threshold, where $\sigma_{c_0}$ can be taken at the inception of linear instability or the initial condition, when, as explained previously, the magnitude of standard deviation is proportional to the strength of the random fluctuation, that is, $\sigma_{c_0} \propto \sqrt{D}$. In the case of linear kinetics with constant exchange current, that is, $R=R_0 \left( \mu_{\text{res}}-\mu \right)$, where $R_0$ is a constant, we have $K = (\mu(c)-\mu(c_0)) R_0/R$. Thus, we obtain the following critical current $R_c$ above which a thermodynamically phase separating system is kinetically stabilized, consistent with Herrmann's conclusion \cite{Herrmann2012}.
\begin{equation}
  \frac{R_c}{R_0} \propto \frac{1}{\ln{1/\sqrt{D}}}.
\end{equation}

\section{Applications} \label{sec:applications}
\subsection{Li-ion Battery Porous electrodes} \label{sec:LFP}

We first look at the example of lithium iron phosphate (LFP,) the canonical and most studied phase-separating solid-state lithium intercalation material. The lithiation reaction is $\text{Li}^+ + \text{e}^- + \text{FePO}_4 \to \text{LiFePO}_4$  and the reverse reaction is called delithiation. The chemical potential of lithium in the solid can be modeled by the regular solution model given by Eq. $\ref{eqn:regular_solution}$ as a function of the lithium fraction in the solid, with $c=0$ being $\text{FePO}_4$ and $c=1$ being $\text{LiFePO}_4$. We model the population dynamics using (a) a classical exchange current, $\sqrt{\left(c(1-c)\right)}$,  (b) a thermodynamically consistent exchange current derived from transition state theory ($(1-c)e^{\mu/2}$) and (c) an experimentally measured exchange current \cite{Lim2016}, all with Butler-Volmer overpotential dependence.

\begin{figure*}[htb]
  \centering
  \includegraphics[width=0.31\textwidth]{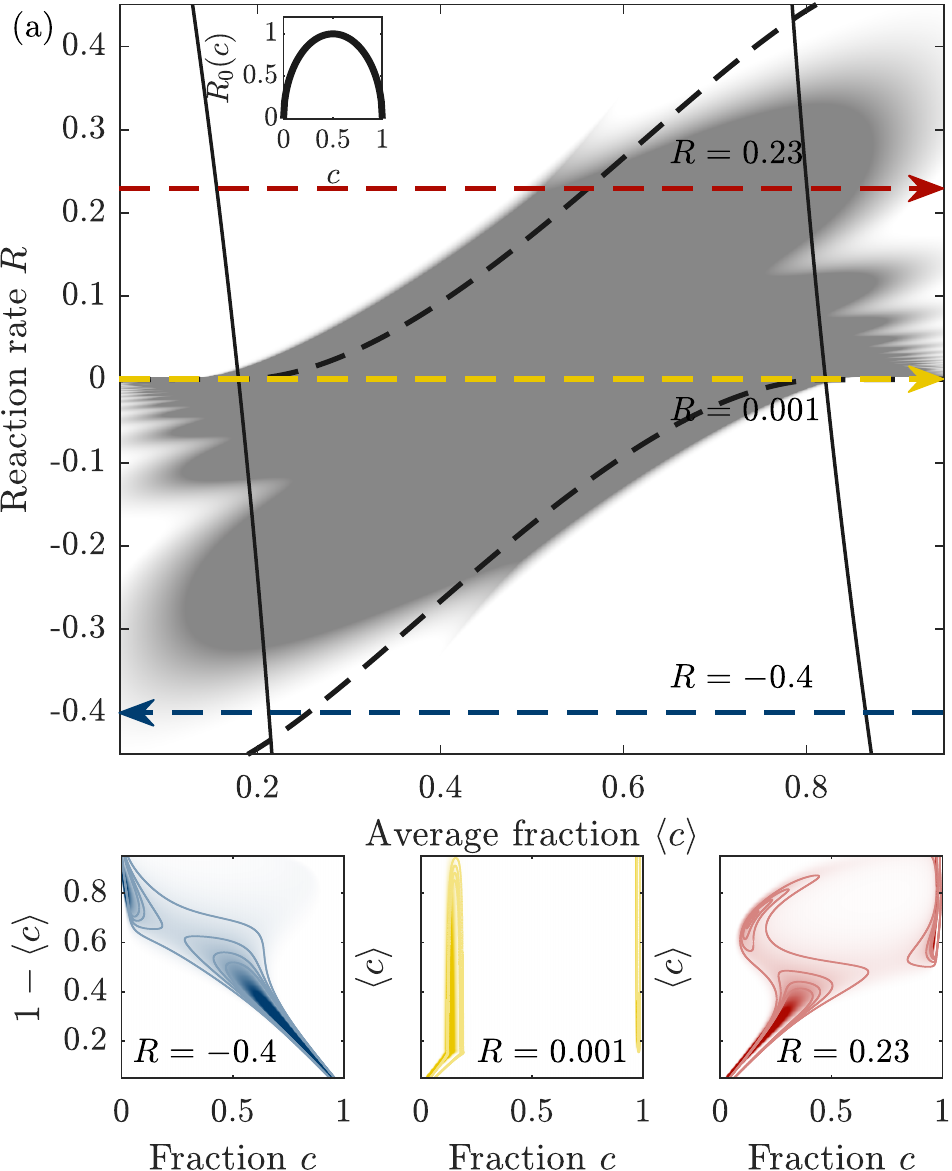} \hspace{0.2cm}
  \includegraphics[width=0.31\textwidth]{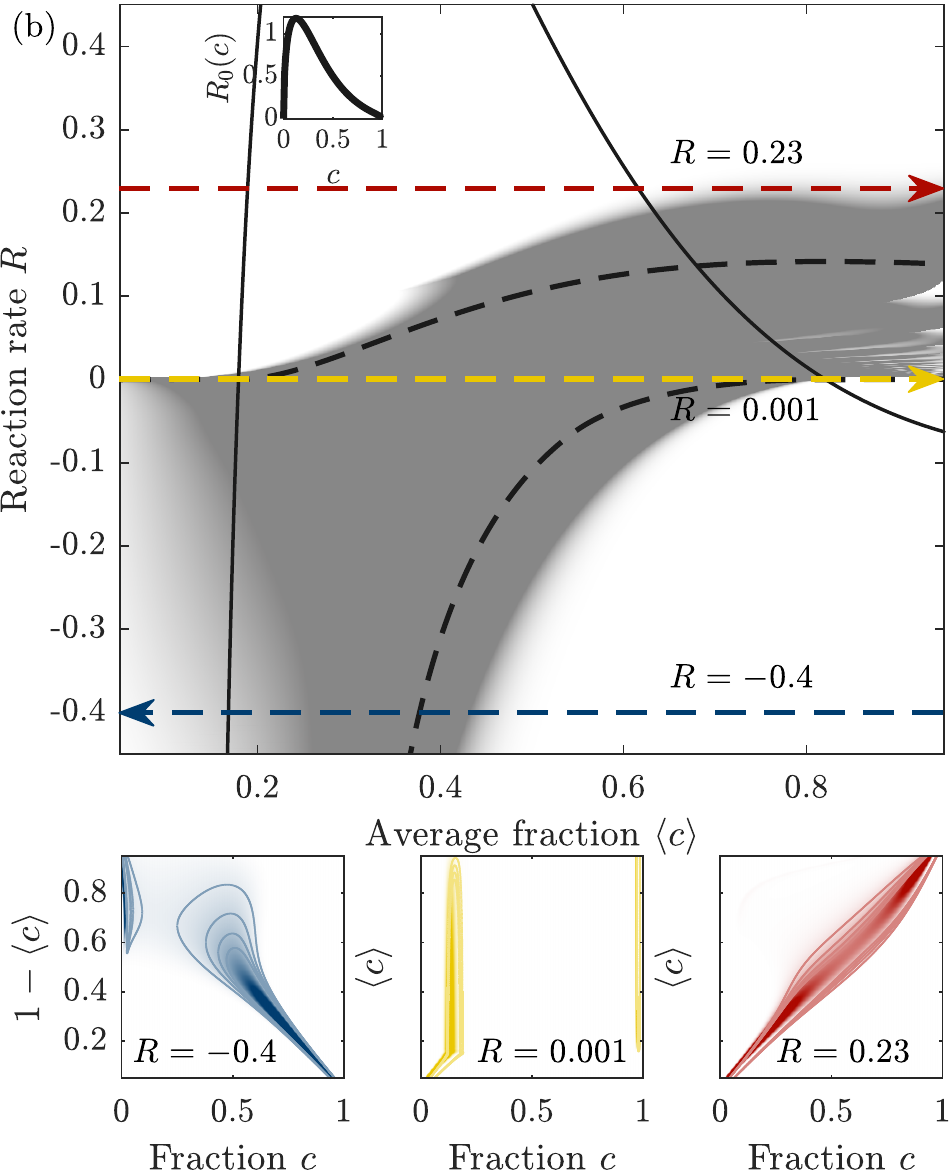} \hspace{0.2cm}
  \includegraphics[width=0.31\textwidth]{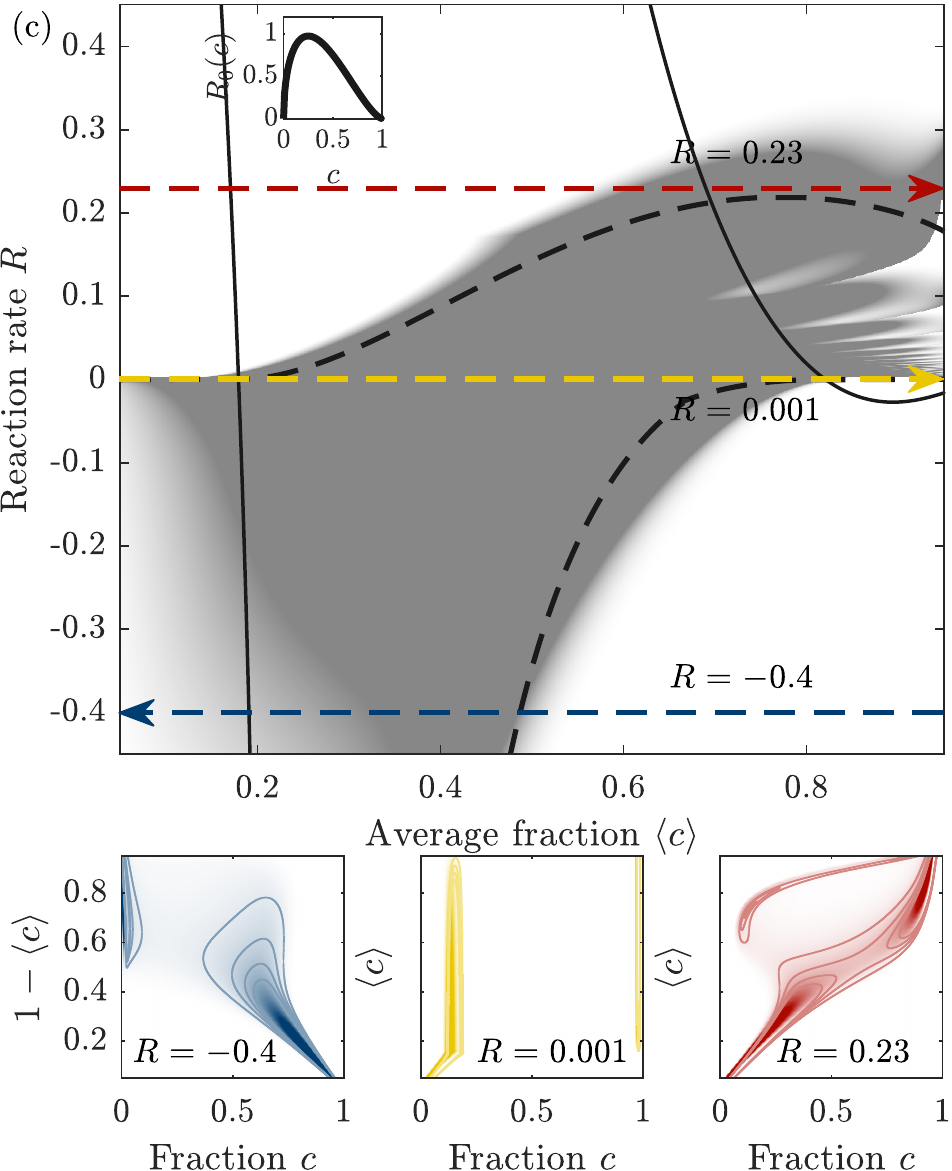}
  \caption{Population dynamics applied to lithium iron phosphate electrode. The main plots show the degree of phase separation as a function of the current applied and average fraction. The prominence of the second peak in $f(t,c)$ is equal to or greater than 1 for the most intense gray color and 0 for white. The plots below show the evolution of the probability distribution in time at a certain constant current whose trajectory is denoted by the same color in the main plot. The value $f(t,c)$ is equal to or greater than 10 at the greatest intensity and 0 for white. The thermodynamics is modeled by regular solution model ($\Omega = 3.4$, $D_0 = 2 \times 10^{-4}$). Insets show the exchange current $R_0(c)$. Within the solid black curves, the system is linearly unstable $\partial R/\partial c>0$. Dashed lines encircle the predicted region of phase separation using Eq. \ref{eqn:PS_criteria}. \label{fig:LFP_PS}}
\end{figure*}

In this section, we focus on the population dynamics of an ensemble of LFP particles under constant current. Fig. \ref{fig:LFP_PS} shows the dynamic phase diagram of the LFP model as a function of current and average Li fraction and the evolution of the probability distribution function in time at select currents. With a positive current (reaction rate), the average Li fraction increases, which corresponds to going horizontally from left to right in the main figure, and vice versa for a negative current. In the previous section, we define phase separation to be when the PDF is bimodal. This is shown by the gray area in the main plot. The color indicates the prominence of the second largest peak, which qualitatively measures the degree of phase separation. To clearly delineate the phase separation region, the color map is at saturation when the value is 1. The point of exit from phase separation is oscillatory with respect to total reaction rate. This is due to the varying frequency observed in the regime of oscillatory phase transition, which is addressed extensively previously and is beyond the scope of this work\cite{Herrmann2012,Herrmann2014,Dreyer2011}.

At low currents, regardless of the exchange current, thermodynamics drives the particle ensemble to separate into Li poor and Li rich populations, which correspond to growing and shrinking peaks seen at $i/i_0 = 10^{-3}$. As shown in the dynamic phase diagram, at low currents (but not $R \to 0$, as explained below), the phase separation initiates at around the nearest spinodal point.

As mentioned above, Herrmann et al. \cite{Herrmann2012,Herrmann2014,Dreyer2011} performed extensive analysis on the various regimes of the population dynamics at different currents for the simple kinetics $R=R_0 \Delta \mu$. Our more general model approaches it asymptotically as $R \to 0$. In the Kramer's regime, the reaction rate is proportional to $\exp{\left(-b/D\right)}$, where $b$ is the free energy barrier, or the difference between the local maximum and minimum of $a-\mu_{\text{res}}c$. Since the reaction rate is equal to the rate of phase transformation in this regime, as $b/D$ increases, the rate of nucleation decreases exponentially. To reach the quasistationary limit, that is, for the phase separating region to span between the two thermodynamic phases (binodal points), the rate must be smaller than the Kramer's rate at the largest possible free energy barrier ($\mu_{\text{res}}=0$), which is around $10^{-422}$ for the particular set of parameters chosen ($\Omega = 3.4$, $D_0 = 2 \times 10^{-4}$), therefore the dynamic phase diagram shown is effectively only for current at or above the Kramer's regime, and the system must at least reach the spinodal point for the nucleation to occur.

The solid curves denote the boundary of linear instability, $s=0$. In between the two curves, $s>0$. At zero current, the region of linear instability is also the spinodal region. The region of phase separation does not follow the shape of the region of linear instability. In fact, nucleation is delayed with increasing current. We seek to quantitatively understand this trend by approximating phase separation to be when the support of the distribution, estimated by the mean plus or minus one standard deviation $\sigma_c = \sigma_{c_0}e^K$, is contained in $[0,1]$, because when phase separation occurs, the new phase is close to $c=1$ for $R>0$ and close to $c=0$ for $R<0$. In other words, we approximate single phase with the following condition,
\begin{equation} \label{eqn:PS_criteria}
  0 < c \pm \sigma_c < 1.
\end{equation}

For a qualitative understanding, we choose an order-of-magnitude estimate for the initial variance, $\sigma_{c_0} = \sqrt{D}$. $c_0$ is the first point where $s=0$ or the initial average concentration used in the simulation (0.05 or 0.95), whichever comes later in the direction of reaction. The predicted region of phase separation is bounded by the dashed lines. This simple approximation captures the general feature of the boundary of phase separation. The form of reaction kinetics, in particular the autocatalytic rate $\partial R/\partial c$, is the dominant factor in determining the phase separating region. It also indicates that the delayed nucleation is mainly due to decreasing $s/R$ with increasing reaction rate.

The regimes studied by Herrmann at higher current no longer hold for certain exchange currents. For the simple kinetics, the distribution stays unimodal above a certain current (around 0.1). The qualitative trend remains true for the exchange current (a). And since exchange current (a) is symmetric with respect to $c=0.5$, the phase behavior is also symmetric between lithiation and delithiation. However, the asymmetric exchange currents (b) and (c) induce autocatalytic effect during delithiation and cause a larger degree of phase separation than lithiation while still suppressing the phase separation above a certain lithiation current. In fact, both (b) and (c) predict that the phase separation persists at large negative current. This demonstrates a new emergent phenomenon that autocatalytic effects add to the population dynamics.

\begin{figure}[htb]
  \includegraphics[width=\columnwidth]{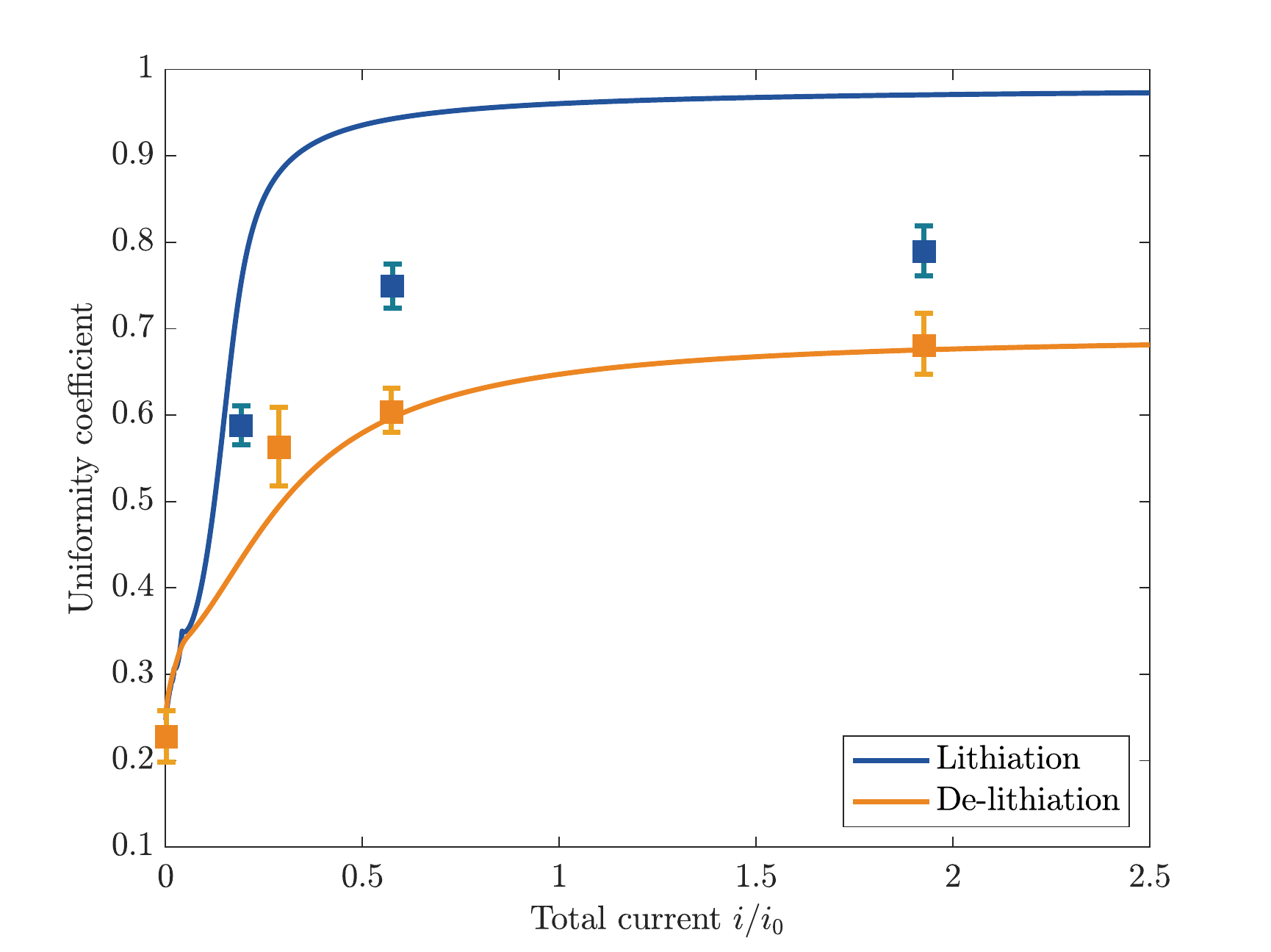}
  \caption{Uniformity coefficient from the model (solid curves) and experiment \cite{Lim2016} (squares with error bars). The model uses the experimental exchange current given in the same reference, $R_0 = 3(1-c)\sqrt{c(1-c)}$.}
  \label{fig:uc}
\end{figure}

Following Lim \cite{Lim2016}, we also define uniformity coefficient, $1- \sigma_c / \sqrt{c(1-c)}$, which is 1 when the probability of finding particles at the average concentration is 1, $P(c=\avg{c})=1$, and 0 when maximum $\sigma_c$ is attained, that is, $P(c=0)=1-\avg{c}$ and $P(c=1)=\avg{c}$. Fig. \ref{fig:uc} shows that the experimental exchange current (c) results in increasing uniformity with increasing total current and that, due to the skewed exchange current and reasons stated above, lithiation results in more uniform distribution than delithiation. The result is in qualitative agreement with the experimental result from Lim \cite{Lim2016}. The theory overpredicts the uniformity for lithiation, which could be attributed to additional noise unaccounted for, such as particle size distribution and spatial inhomogeneity of the reaction kinetics. The asymmetry is also observed in another study by Li et al. \cite{Li2014} where the active particle fraction increases as a function of current and it is higher for lithiation than for delithiation. This is consistent with the asymmetric exchange current since a smaller active particle fraction indicates a higher degree of phase separation and a lower uniformity coefficient.

\subsection{Driven phase separation at the critical point}
In this section, we present an example of a thermodynamically single-phase system that undergoes fictitious phase separation in certain reaction conditions. We choose the regular solution model with $\Omega=2$, so that the system is at the critical point. As mentioned earlier, $\partial R / \partial c = -k_0 \partial \mu / \partial c + \left( \mu_{\text{res}} - \mu \right) \partial k/\partial c$. At the critical point $c_0$, $\mu'(c_0)=0$, $\partial R / \partial c(c_0) = \left( \mu_{\text{res}} - \mu(c_0) \right) \partial k/\partial c(c_0)$, whether the reaction is autocatalytic or autoinhibitory is solely determined by the explicit dependence of reaction rate on concentration $k(c)$ and the direction of reaction.

For illustration purposes, we choose Butler-Volmer kinetics with the exchange current, $R_0 = (1-c)^2 e^{\alpha \mu}$ \cite{Bazant2013}. In a lattice model, the exponent 2 indicates that the transition state excludes 2 empty sites \cite{Bazant2013}. The quadratic dependence on $(1-c)$ enhances the autocatalysis and accelerates the backward reaction as $c$ decreases. Fig. \ref{fig:demo_autocat} is a simulation with such thermodynamic and kinetic models, at $\mu_{\text{res}}=-0.2$ and total reaction rate $-0.1$. It illustrates the diverging characteristics during the backward reaction.

\begin{figure}[htb]
  \centering
  \includegraphics[width=\columnwidth]{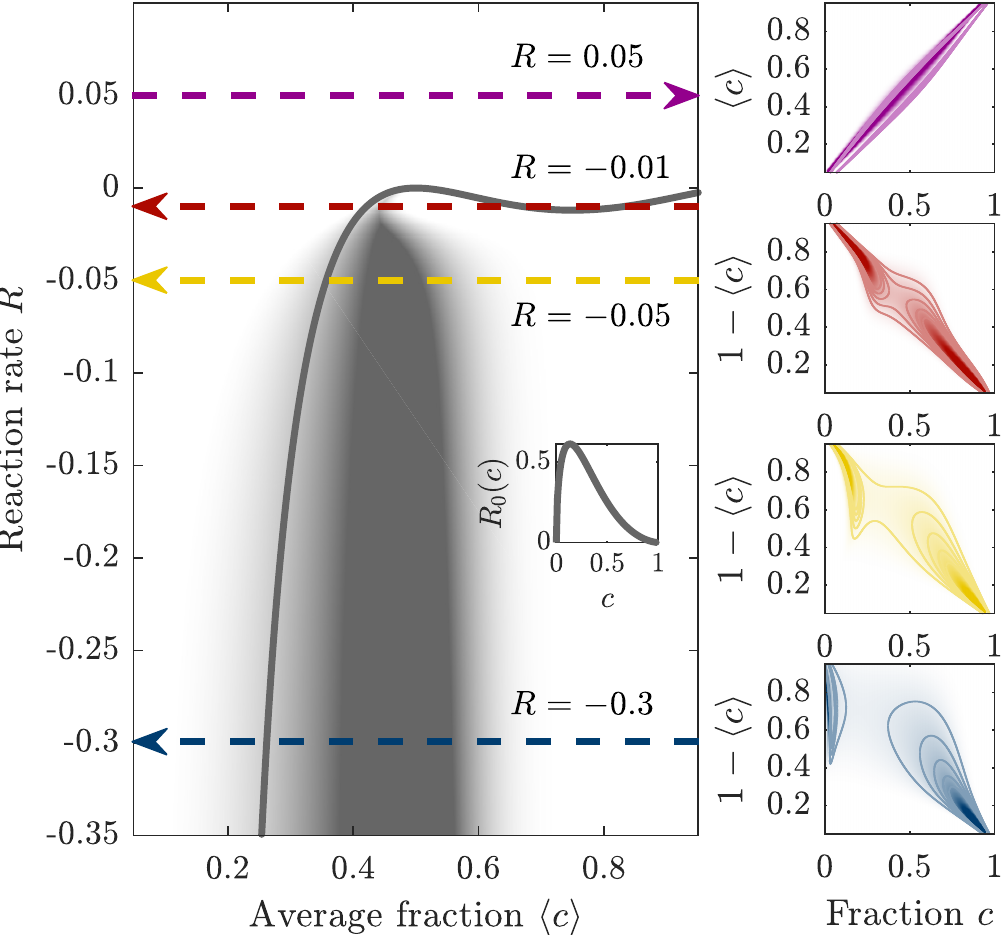}
  \caption{Population dynamics applied to a system at the critical point ($\Omega=2$, $D_0 = \times 10^{-3}$). The exchange current is shown in the inset. The gray area is the region of fictitious phase separation. The colormap is the same as that of Fig. \ref{fig:LFP_PS}. The system is linearly unstable below the gray solid curve. \label{fig:NMC_PS}}
\end{figure}

Fig. \ref{fig:NMC_PS} shows the kinetic phase diagram of the system. Instability and bimodal distribution occur with very small negative reaction rate. Unlike LFP, increasing backward reaction rate leads to the expansion of the phase separation region. During the backward reaction, the distribution diverges so much that the population separates into two distinct ``fictitious phases'' while the forward reaction shows stable single-phase behavior. The region of phase separation is even more asymmetric than LFP with the asymmetric exchange current due to the second order dependence on $1-c$. The fictitious phase separation region is contained within the linearly unstable region, which is below the solid gray curve. We do not give the approximated phase separation region like the one above for LFP, because when the second peak in the PDF appears, the peak position is not always close to 0 or 1.

This leads to important applications in the interpretation of in-situ and operando X-ray diffraction (XRD) experiments of electrodes, which measures the population density as a function of lattice parameters and thus the state of charge of the particles\cite{Zhou2016,Liu2014,Singer2014,Hong}. One should be cautious in deriving phases of a material from the XRD spectrum of a dynamical experiment such as charge and discharge, as the XRD peaks may arise as a result of kinetic effects.

The phase separation defined for an particle ensemble can also inform us about the initiation/nucleation of phase separation within particles. In particular, the example presented here suggests that pattern formation may also arise when driven away from equilibrium through autocatalytic reactions even though the reactive mixture is thermodynamically stable. This could have implications in understanding and engineering phase separation in a variety of systems and patterns driven by electron transfer reactions. Negative differential resistance ($\partial R/\partial \Delta \mu<0$) is not studied here in the context of population dynamics but can also lead to fictitious phase separation.

\section{Summary}

We lay out a general theory of describing the equilibrium and dynamics of an ensemble of spatially homogeneous reactive particles that undergoes material exchange with the reservoir. The theory extends beyond the discrete particle simulation or the simplified reaction kinetics that are studied previously. Under the reaction-controlled condition, reaction kinetics determines the characteristics of the governing equation for the probability distribution of the particle ensemble. In regions of autocatalysis or autoinhibition, the probability distribution exhibits shock wave behaviors. We established the framework of introducing mole fluctuation for any generalized reaction kinetics in a thermodynamically consistent way. The corresponding Fokker-Planck equation gives the equilibrium probability distribution that is consistent with statistical physics.

Through analytical approximation and numerical analysis of several model systems, we studied how reaction kinetics of the system affects the population dynamics and in some cases transforms the entire landscape of phase behavior for systems with the same thermodynamic properties. Suppression of phase separation or the induction thereof in single-phased systems have been observed. The analytical approximation for the boundary of phase separation based on the integral of autocatalytic rate shows good agreement with numerical simulation. This reinforces our understanding that autocatalytic effects lie at the core of such phase separation behavior.

Therefore, knowledge of the reaction kinetics is crucial in understanding the dynamics of reaction-controlled systems. One can also envision engineering reactive surfaces to achieve desirable kinetic properties. The emergent phenomena observed here motivate us to further explore kinetically controlled phase separation in experiments using in-situ analysis such as the aforementioned X-ray diffraction.

Although we did not address any possible spatial correlation or dependency in this paper, the theory can be extended to incorporate the external environment that the particles are subjected to by making the probability space a Cartesian product of the particle state space and physical space. For example, we can introduce other fields such as mass, temperature and electric potential to consider transport phenomena, as well as the particle ensemble's interaction with the external fields. Particles in a porous medium can also be influenced by its highly localized environment. Any intrinsically variable property or such local variability that follow certain probability measures can also be incorporated in the theory. The population dynamics can also shed light on the spatially dependent behavior within the particles themselves. For example, by switching from concentration field in a physical space to probability distribution, population dynamics offers a spatially-agnostic statistical description of the concentration field such as its mean and the variance for as long as it maintains isotropy and before patterns with sharp concentration gradient take over, therefore potentially serving as a precursory analysis for the inception or even termination of pattern formation.

\begin{acknowledgments}
This work was supported by Toyota Research Institute through the D3BATT Center on Data-Driven-Design of Rechargeable Batteries.
\end{acknowledgments}

\bibliography{Zhao}

\end{document}